\documentclass[a4paper,11pt]{article}
\usepackage{jinstpub} 
\usepackage{comment}

\title{\boldmath Impact of angle-dependent recombination on neutrino energy reconstruction in LArTPCs}

\author[a,1]{J. Vences,\note{Corresponding author.}}
\author[a,b]{J. Maneira}
\author[a]{C. Vilela}
\affiliation[a]{Laborat\'{o}rio de Instrumenta\c{c}\~{a}o e  F\'{\i}sica Experimental de Part\'{\i}culas - LIP, 1649-003 Lisboa, Portugal}
\affiliation[b]{Faculdade de Ci\^{e}ncias - FCUL, Universidade de Lisboa, 1749-016 Lisboa, Portugal}

\emailAdd{joanavences@lip.pt}

\abstract{The technology of Liquid Argon Time-Projection Chambers (LArTPC) plays a very important role in modern neutrino physics. It allows precise tracking and calorimetry in detectors with very large volume and mass. One important component of LArTPC calorimetric energy reconstruction is the electron-ion recombination effect.
Existing experiments have provided constraints to its empirical description,  including an explicit dependence on the particle's angle with respect to the TPC's electric field. 
In this paper we analyze the impact of angle-dependent recombination models on the reconstruction of neutrino energy in LArTPCs. We examine in detail the predictions of the existing models, showing that there is still considerable uncertainty for the angular dependence prediction at the low values of $dE/dx$ typical of minimum ionizing particles. We apply \textcolor{black}{adjustments} to overcome unphysical predictions in the non-constrained energy loss regions and evaluate the consequences of that in the bias and linearity of electron neutrino energy responses, using generator-level and electron shower simulations for neutrino energies in the range of 0.5-5 GeV. Comparing prediction with and without angular-dependence, we observe an average energy bias between -0.9\% at 0.5 GeV and -0.6\% at 5 GeV \textcolor{black}{in the reconstructed electron energy}. While small, these values are close to the typical requirement of 2\% for the energy scale uncertainty of future experiments, and indicate the need to obtain actual experimental constraints at low $dE/dx$ and take the effect into account in order to achieve the best precision in future LAr TPC electron neutrino calorimetry.

}


\begin{document}
\maketitle
\flushbottom

\section{Introduction}
\label{sec:intro}
The Liquid Argon Time-Projection Chamber (LArTPC) is a sophisticated detector technology first proposed in the mid 1970's~\cite{Rubbia:1977zz} that has become central to modern neutrino physics. The technology uses the ionization properties of liquid argon, combined with the transport of charges over large distances under a strong electric field, to enable high-resolution 3D imaging of charged particle trajectories as well as precise calorimetry. The time-projection technique allows the number of readout channels to scale roughly with the surface area (or length) of the detector, as opposed to the volume. Given the relatively high density of liquid argon, this makes the technology scalable to multi-kiloton masses. The combination of large detector mass, particle identification and energy reconstruction capabilities, makes LArTPC detectors ideal for neutrino oscillation experiments. 

LArTPCs with several hundred kilograms of argon mass have been deployed in neutrino oscillation experiments, including ICARUS~\cite{ICARUS:2004wqc}, MicroBooNE~\cite{MicroBooNE:2016pwy}, and SBND~\cite{SBND:2025lha}. This technology will be scaled up by more than one order of magnitude for the next-generation DUNE long-baseline experiment~\cite{DUNE:2020lwj}, with each of its far detector modules containing 17 kiloton of argon. The statistical precision given by the large number of neutrino interaction events expected in current and planned LArTPC experiments needs to be matched by a precise characterization of the detector response to neutrino interactions. In particular, the energy scale uncertainty of MicroBooNE is at the 2\% level~\cite{MicroBooNE:2019efx}, and this is also the target for the DUNE experiment~\cite{DUNElbl}.

One important component of LArTPC calorimetric energy reconstruction is the electron-ion recombination effect, which needs to be taken into account to avoid bias. This effect has been studied extensively~\cite{jaffe,onsager,box,icarus_recom}, and several models capture its dependence on the ionisation density, typically expressed in terms of energy loss per unit length ($dE/dx$), and on the strength of the electric field applied to the liquid argon. A few studies have also characterized the dependence of electron-ion recombination on the angle between the ionizing particle trajectory and the electric field direction~\cite{argoneut_recom,icarus_angular}.

In this paper we analyze the impact of angle-dependent recombination models on the reconstruction of neutrino energy in LArTPCs\textcolor{black}{, focusing specifically on electrons}. The paper is organized as follows: in Section~\ref{sec:data} we describe the simulated data used for this publication and in Section~\ref{sec:erec}, we outline the calorimetric approach to neutrino energy reconstruction and introduce the relevant kinematics for our analysis. We discuss angle-dependent recombination models in Section~\ref{sec:models} and in Section~\ref{sec:impact} we study the impact of these models on neutrino energy reconstruction \textcolor{black}{by analysing the effect on the electron energy (the effect on hadronic energy will not be evaluated in this paper)}. We summarise the conclusions of this study in Section~\ref{sec:end}.

\section{Simulated data sets}
\label{sec:data}
Two sets of simulated data have been used in this analysis. To study the kinematics of particles produced in neutrino-Ar interactions a set of one million $\nu_e$ events was produced with version 2.12.10 of the \textsc{Genie} generator~\cite{GENIE} using the \textsc{Nuisance} framework~\cite{Stowell:2016jfr}. The events in this sample are distributed roughly uniformly in neutrino energy up to 10~GeV, so that the kinematics of outgoing particles can be analyzed as a function of the interacting neutrino energy with no assumptions on beamlines or other experimental conditions. These events are not processed through a detector simulation and only generator-level information is used. All interaction  modes in the \textsc{Genie} default tune were included in the simulation.

Our analysis also includes an inspection of the structure of ionizing tracks in electromagnetic showers induced by $\nu_e$ charged current (CC) interactions on argon. For this purpose, we use ten samples of \textsc{Genie} events generated at fixed energies, from 0.5~GeV to 5.0~GeV in steps of 0.5~GeV. These events are then processed with \textsc{Edep-sim}~\cite{edep-sim}, an energy deposition simulation based on \textsc{Geant4}~\cite{ALLISON2016186,1610988,AGOSTINELLI2003250}. To avoid biasing our results towards a particular detector geometry, the neutrino-Ar events are simulated in a large volume consisting only of liquid argon.

\section{Neutrino energy reconstruction in LArTPCs}
\label{sec:erec}

Neutrino energy is reconstructed from measurements of charged particles resulting from its interactions. CC neutrino interactions are crucial since they provide information on the flavour and energy of the interacting neutrino. From CC neutrino interactions results a leptonic and hadronic component from which the neutrino energy is reconstructed:
\begin{equation}
\label{eq:1.1}
E_\nu = E_{lep} + E_{had}
\end{equation}

\begin{figure}[htbp]
\centering
\includegraphics[width=.3\textwidth]{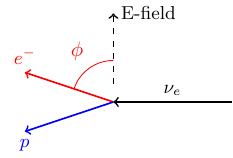}
\qquad
\includegraphics[width=.55\textwidth]{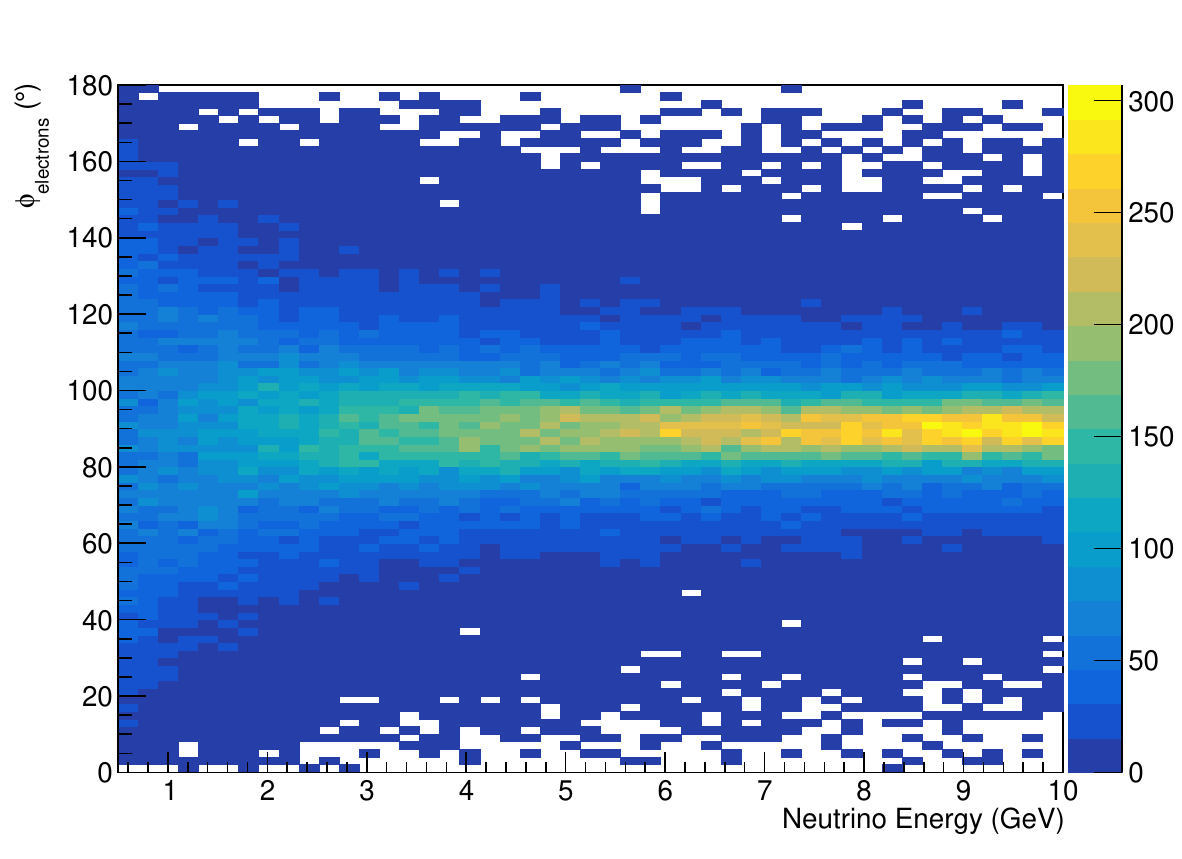}
\caption{(Left) Example of a CC neutrino interaction considered in the data set. The direction of the neutrino is assumed to be perpendicular to the electric field. (Right) Angle between electrons resulting from a CC neutrino interaction and the electric field ($\phi$) versus the neutrino energy. \label{fig:ang2d}}
\end{figure}

For the purposes of this study, the first set of simulated data as described in Section \ref{sec:data} was used, produced with the \textsc{Genie} generator. Out of the one million events in the data set, only CC interactions were considered. The range of neutrino energy considered was from 0.5 GeV to 10 GeV. We assume the neutrino beam to come from a fixed direction with the electric field perpendicular to it, as shown on the left side of Figure \ref{fig:ang2d}. \textcolor{black}{It should be noted that even for short baseline experiments, the neutrino angular spread is much smaller than the dependence of the lepton angle on neutrino energy shown in Fig \ref{fig:ang2d}. Therefore, the assumption that the neutrino beam comes from a fixed direction does not significantly impact the applicability of our results.} 

In order to analyse the impact of electron-ion recombination on the reconstruction of neutrino-Ar interactions, it is relevant to consider the angle between charged particles resulting from the interaction and the electric field, $\phi$. We are interested in studying how the distribution of the angle varies with the neutrino energy. Shown in the right side of Figure \ref{fig:ang2d} is the distribution of the angle $\phi$ for electrons versus the neutrino energy.

At higher energies, the electrons will follow the original direction of the neutrino closely and as such, the distribution is centered around $\phi = 90^\circ$. For lower neutrino energies, the angle distribution is quite spread out. This dependence of $\phi$ with the neutrino energy is expected due to kinematics.

\begin{figure}[htbp]
\centering
\includegraphics[width=.6\textwidth]{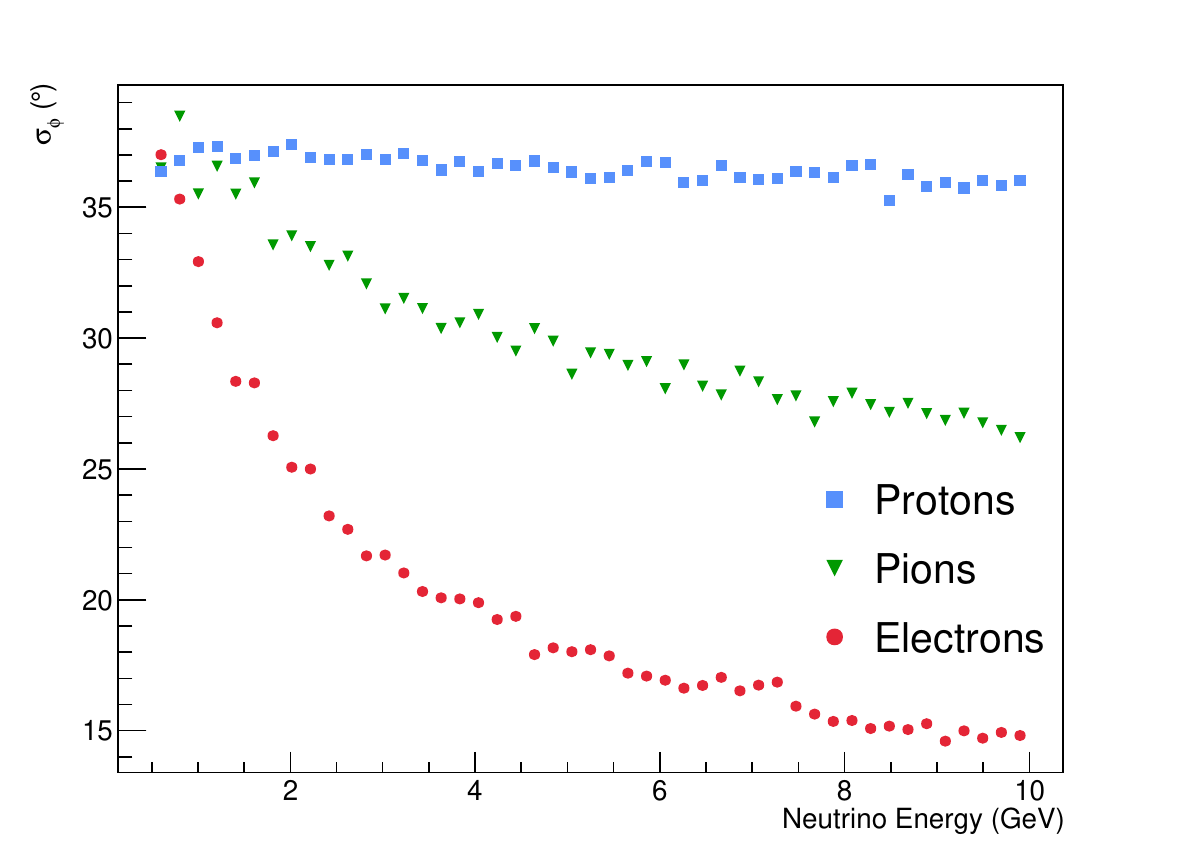}
\caption{Standard deviation of the angle between various particles resultant of CC neutrino interactions and the electric field as a function of the neutrino energy.\label{fig:std}}
\end{figure}

The same relationship was studied for protons and pions. In Figure \ref{fig:std} is shown the standard deviation of $\phi$ ($\sigma_\phi$) for protons, electrons and pions as a function of the neutrino energy.
The kinematics constraints are different for protons and $\sigma_\phi$ remains fairly constant throughout the neutrino energy range. As for electrons and pions, due to their lower masses, the standard deviation does vary with the neutrino energy. For electrons, the standard deviation is around $35^\circ$ for a neutrino energy of 0.5 GeV and decreases to around $15^\circ$ for a 10 GeV neutrino.

\begin{figure}[htbp]
\centering
\includegraphics[width=.6\textwidth]{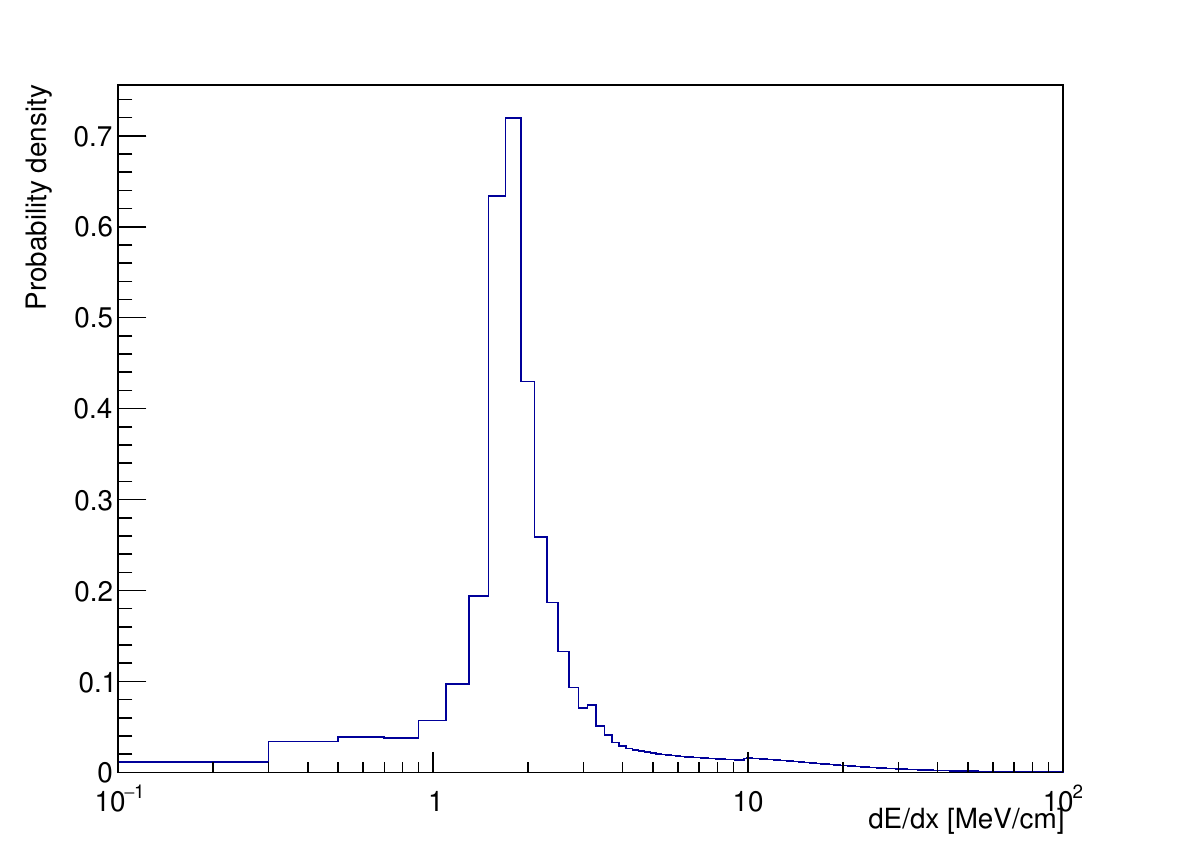}
\caption{Distribution of the stopping power for the electromagnetic shower segments.\label{fig:dedx}}
\end{figure}

Due to the fact that protons have a larger  $dE/dx$, they more easily ionize the liquid argon and the recombination effect is more significant. However, we focus instead on the electron, which carries a larger fraction of the neutrino energy and whose angular distribution depends more strongly on the neutrino energy. At the GeV scale, the electrons produced in $\nu_e$ CC interactions are typically above the critical energy in argon, 45~MeV~\cite{NIST}, and therefore they induce electromagnetic showers. The detailed simulation of energy deposition with \textsc{Edep-sim} is used to study the recombination effect on each electron track segment that forms the shower. The distribution of track segment $dE/dx$ for electromagnetic showers ranging from 0.5~GeV to 5~GeV is shown in Figure~\ref{fig:dedx}, where it can be seen that a significant number of segments has a low $dE/dx$, similar to a minimum ionizing particle.

\textcolor{black}{The choice of Geant4 step size and tracking thresholds may affect the distribution of $dE/dx$ in electromagnetic shower segments shown in Fig.~\ref{fig:dedx}. We note that the default configuration of \textsc{Edep-sim} was used in this work, which is also used for the simulation of the DUNE near detector~\cite{DUNE:2021tad}. Therefore, the results are directly applicable to the existing DUNE near detector studies. We leave systematic studies of the impact of Geant4 step size and tracking threshold for future work.
}

\textcolor{black}{\section{Angular-dependent recombination models}\label{sec:models}}
Charged particles traversing through liquid argon will ionize the argon atoms, creating electrons and positive ions, which drift in the presence of an electric field. In the process known as recombination, ions drift much more slowly than electrons, increasing the probability that electrons are captured by them. The argon ions have a much smaller drift velocity and as such, will stay closer to the production zone for a longer time~\cite{lar-RECOM-prop}. Recombination typically happens after electrons thermalize and are carried by the electric field through the ion-dense regions. 

The Onsager theory~\cite{onsager} describes an electron captured by its parent ion. This initial recombination effect has been shown to be negligible~\cite{argoneut_recom}. In the case of the Jaffé theory~\cite{jaffe}, ionization charge is distributed around a charged particle track in the form of a column in which the distributions are gaussian. The probability of an electron undergoing recombination is dependent on the charge density and the electric field. Another way to model recombination is to consider electron and ion mobility to be negligible. In this model, developed by Thomas and Imel~\cite{box}, the ionization charge is distributed uniformly in a cubic box. 

Following closely the approach in Ref.~\cite{icarus_angular}, we consider two parameterizations of the recombination process in LArTPCs. The Birks equation, which is an approximation of the Jaffé theory, is given by:
\begin{equation}
    \label{eq:4.3}
    \frac{dQ}{dx} = \frac{A_B}{W_{ion}}\frac{dE/dx}{ 1 + \frac{k_B}{\epsilon \rho}f(\phi) dE/dx}
\end{equation}
where $dQ/dx$ is the linear density of charges surviving the electron-ion recombination process, \mbox{$W_{ion} = 23.6$~eV/electron} is the average energy needed to ionize an argon atom, $\rho$ is the liquid argon density, $\epsilon$ is the electric field, $A_B$ and $k_B$ are free parameters of the model, and $f(\phi)$ encodes the angular dependence of the model. In the case of no angular dependence, $f(\phi) = 1$.

The modified box model, based on the model mentioned previously by Thomas and Imel, is given by the following equation: 
\begin{equation}
    \label{eq:4.4}
    \frac{dQ}{dx} = \frac{\log(\alpha + \mathfrak{B} dE/dx)}{\mathfrak{B}W_{ion}}
\end{equation}
\textcolor{black}{where $\mathfrak{B} = \frac{\beta_{90}}{\epsilon \rho}f(\phi)$ and} $\alpha$ and $\beta_{90}$ are free parameters of the model.

Both these models present their own issues, being limited in describing the recombination effect consistently in certain regimes. In the case of the Birks model, there are technical difficulties in implementing it in high ionization regions~\cite{argoneut_recom}. On the other hand, the box model performs poorly for low values of $dE/dx$. The modified box model~\cite{argoneut_recom} was developed to correct this issue by allowing $\alpha < 1$ and adjusting the parameter $\beta$.

An explicit dependence on the angle between the charged particle direction and the electric field can be introduced in the models above via $f(\phi)$. This angular dependence can introduce a bias in the charged particle energy reconstruction and potentially create a distortion in the neutrino energy reconstruction. The angular dependence of recombination will depend on the characteristics of the distribution of the initial electron-ion cloud such as its shape and direction~\cite{ellipsoidal}.

In the case of the Birks model, it is generally understood that an angular dependence can be introduced assuming the ionization charge is initially distributed in the form of an infinitely long column around the direction of the ionization track~\cite{jaffe} by setting the angular function: 
\begin{equation}
    f(\phi) = 1/\sin{\phi}.
\end{equation}

As for the modified box model, in Ref.~\cite{icarus_angular}, in addition to the columnar model for the ionization region shape, an ellipsoidal model~\cite{ellipsoidal} is also considered: 
\begin{equation}
    f(\phi) = 1/\sqrt{\sin^2{\phi} + \cos^2{\phi}/R^2},
\end{equation}
where R is a parameter related to the eccentricity of the ellipse. The choice between these shapes impacts the angular dependence of the model. 

The ellipsoidal model for the angular dependence includes both the columnar cloud (for $R \rightarrow \infty$) and the spherically symmetric cloud (for $R = 1$). For both the columnar and ellipsoidal models of the ion-dense region shape, the probability of recombination is smaller at large angles (close to $\phi = 90^\circ$). For smaller angles, both predict a larger recombination effect, i.e., higher loss of drifting electrons, due to a larger overlap between the regions with high electron and ion densities.

\begin{figure}[htbp]
\centering
\includegraphics[width=.6\textwidth]{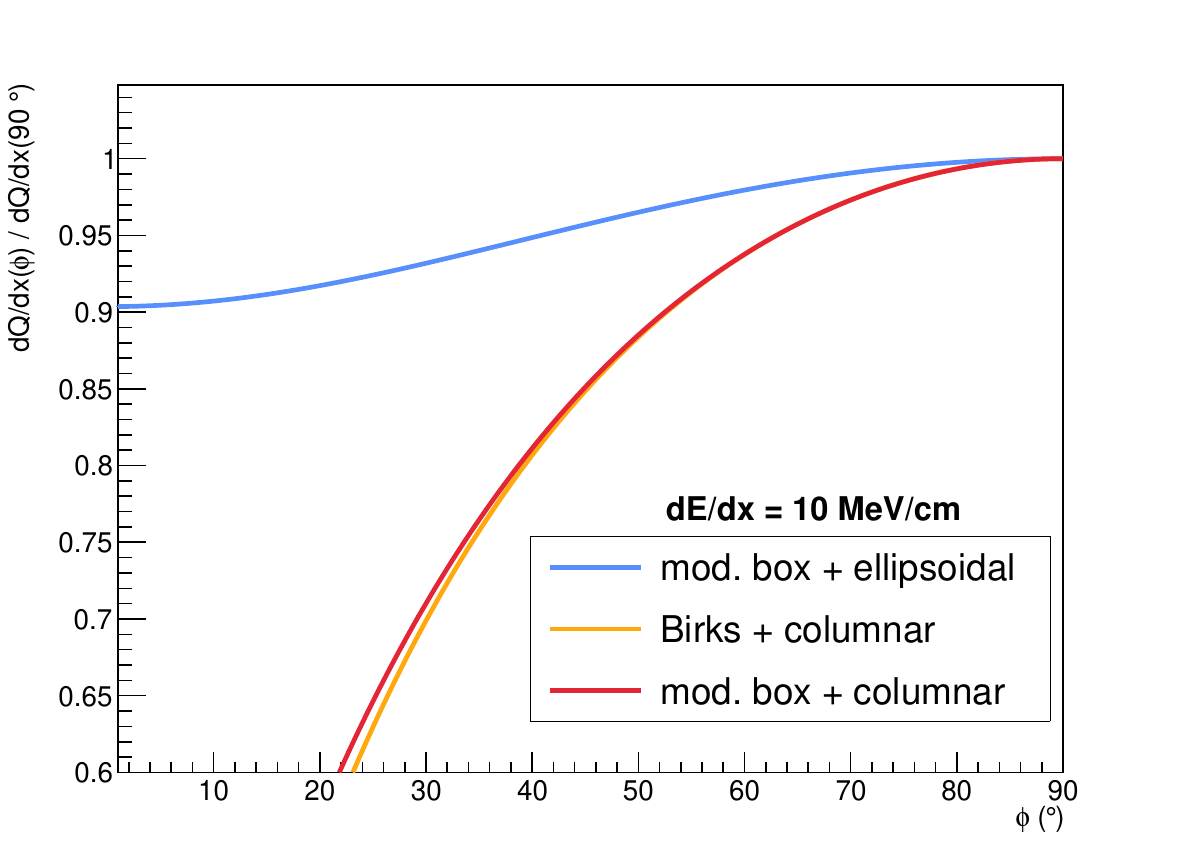}
\caption{Ratio between $dQ/dx(\phi)$ and $dQ/dx(90^\circ)$ for the modified box model with two models of angular dependence (columnar and ellipsoidal)~\cite{icarus_angular} and for the angular-dependent Birks model where the values for the parameters $A_B$ and $k_B$ were taken from \cite{argoneut_recom}.\label{fig:models}}
\end{figure}

The goal of this study is to quantify the angular dependent recombination effect and study its impact on neutrino energy reconstruction. For this purpose, the ratio between $dQ/dx(\phi)$ and $dQ/dx(90^\circ)$ (where the recombination effect is minimal) was calculated for the three models described above (Birks with an angular dependence, columnar modified box model and ellipsoidal modified box model) as shown in Figure~\ref{fig:models}. 

The values used for $\alpha$, $\beta_{90}$ and $R$ were obtained from the ICARUS measurement~\cite{icarus_angular}, in which the ellipsoidal model was identified as the one that best fit their experimental data. For the Birks equation, the values used for the parameters $A_B$ and $k_B$ were measured by the ArgoNeuT experiment~\cite{argoneut_recom}. 
\begin{table}[htbp]
\centering
\caption{Fitted parameters obtained by ICARUS~\cite{icarus_angular} for the modified box model and by ArgoNeuT~\cite{argoneut_recom} for the Birks model. In ICARUS, $\epsilon = 0.492$~kV/cm and in ArgoNeuT, $\epsilon = 0.481$~kV/cm. Both measurements were made for proton data, with the ICARUS measurement for $5 < dE/dx < 12$  MeV/cm and the ArgoNeuT measurement for $5 < dE/dx < 24$ MeV/cm. \label{tab:param}}
\smallskip
\begin{tabular}{lcc}
\cline{2-3}
                                  & ICARUS & ArgoNeuT \\ \hline
\multicolumn{1}{l|}{$A_B$}        &    -    &   $0.80 \pm 0.01$       \\
\multicolumn{1}{l|}{$k_B (\frac{kVg/cm^3}{\text{MeV}})$}        &     -   &    $0.052 \pm 0.001$    \\
\multicolumn{1}{l|}{$\alpha$}     &   $0.904 \pm 0.008$     &   -       \\
\multicolumn{1}{l|}{$\beta_{90} (\frac{kVg/cm^3}{\text{MeV}})$} &   $0.204 \pm 0.008$      &   -       \\
\multicolumn{1}{l|}{$R$}          &   $1.25 \pm 0.02$     & -        
\end{tabular}
\end{table}
The values of these parameters are presented in Table~\ref{tab:param}. \textcolor{black}{Due to the lack of data for electrons and low $dE/dx$, we use the values of the parameters that were obtained with proton data and for $dE/dx>5$ MeV/cm. While this is an extrapolation, the parameters should be particle-independent since the dependence on the particle of the recombination models is fully encoded into the $dE/dx$.} The behaviour of the Birks model and the columnar modified box model are very similar at high values of $dE/dx$. The angular effect is lessened when considering the ellipsoidal model for the ionization region. Even so, for smaller values of $\phi$, the ratio is close to $0.9$. From this plot it can be noted that including an angular dependence in the recombination effect leads to an angular dependence in $dQ/dx$ which has a significant impact on the collected charge. 

\begin{figure}[htbp]
\centering
\includegraphics[trim={1cm 0 1cm 0}, width=.49\textwidth]{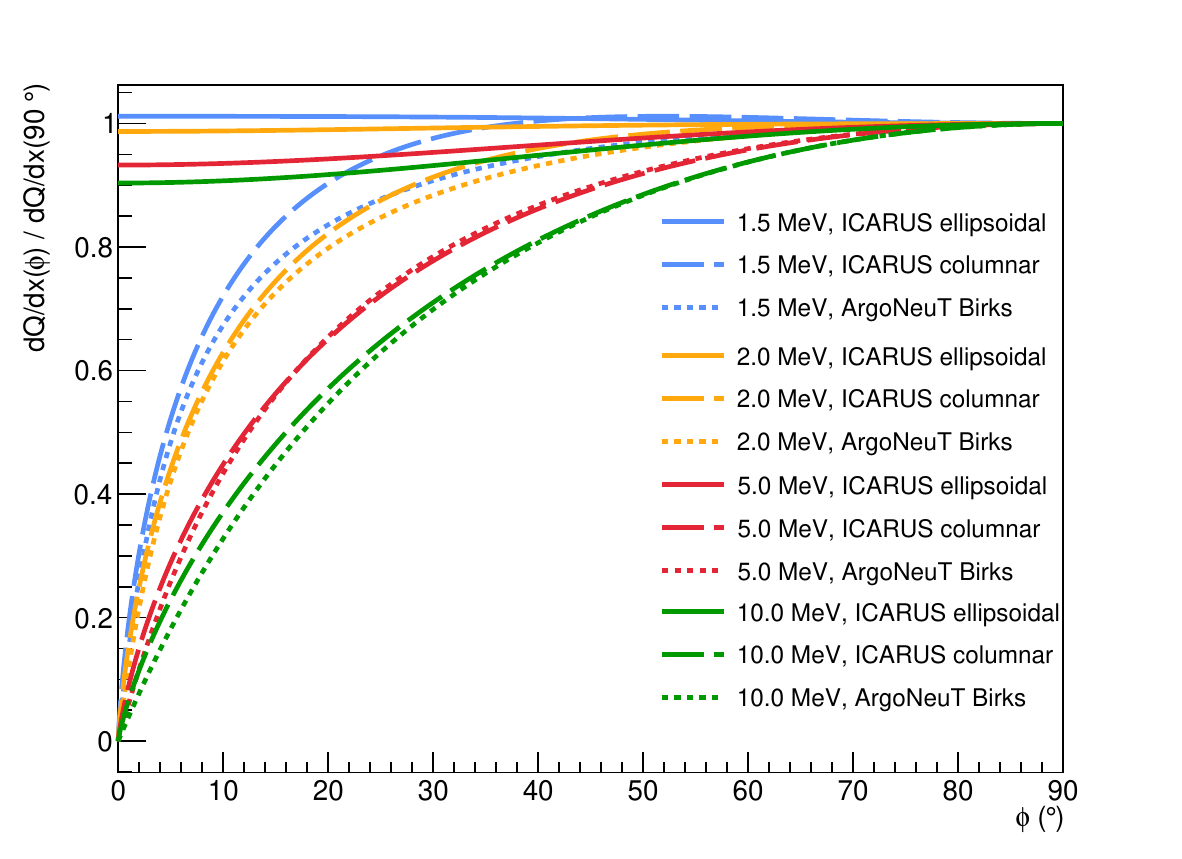}
\includegraphics[trim={1cm 0 1cm 0}, width=.49\textwidth]{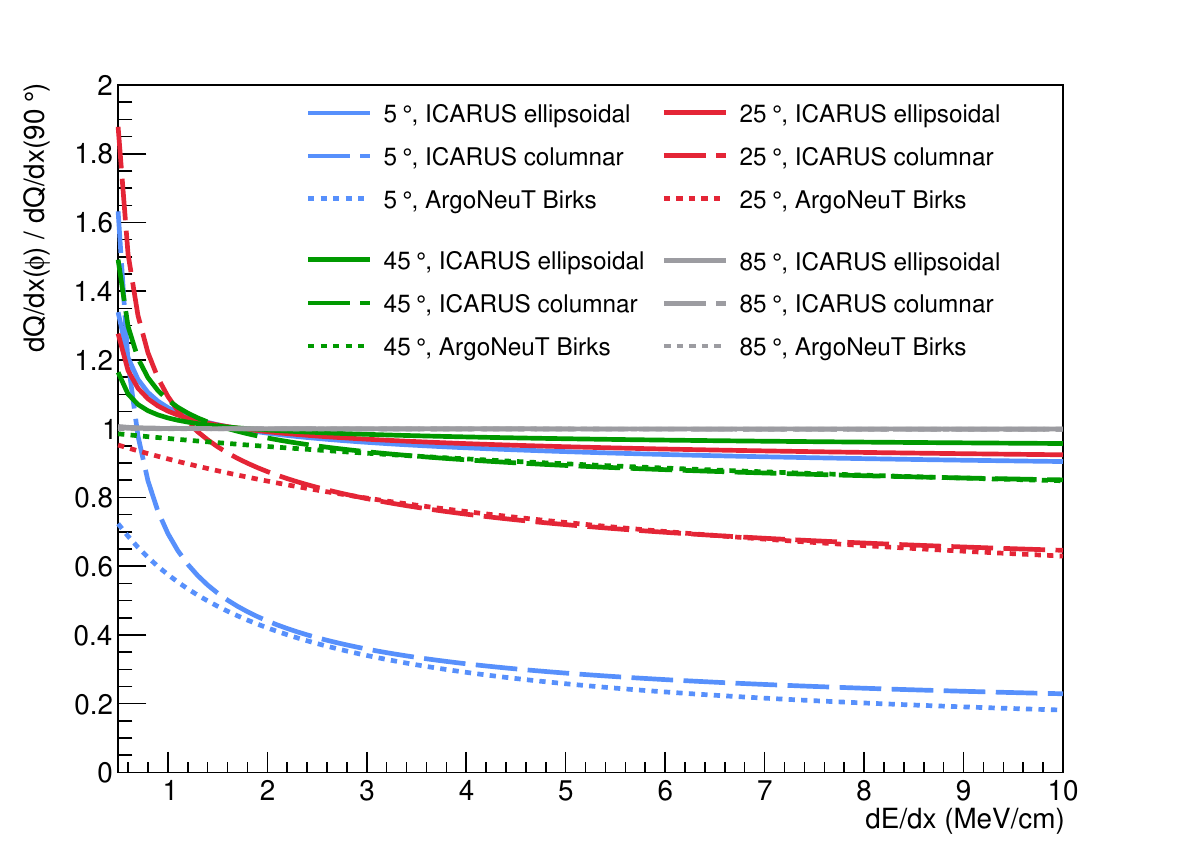}
\caption{(Left) Ratio between $dQ/dx(\phi)$ and $dQ/dx(90^{\circ})$ as a function of $\phi$, the angle between a particle and the electric field for different values of $dE/dx$. The blue curves correspond to the ellipsoidal modified box model. The yellow curves correspond to the angular-dependent Birks law with the values for the parameters published by the ArgoNeuT experiment~\cite{argoneut_recom}. The red curves correspond to the columnar modified box model. For the modified box models the values of the parameters published by the ICARUS experiment \cite{icarus_angular} were used. (Right) Ratio between $dQ/dx (\phi)$ and $dQ/dx(90^{\circ})$ as a function of $dE/dx$ for different values of $\phi$. The solid lines correspond to the ellipsoidal modified box model. The dotted lines correspond to the angular-dependent Birks law. The dashed lines correspond to the columnar modified box model.\label{fig:dqdx_ratio}}
\end{figure}

Since this study is focused on electrons resulting from neutrino interactions and their induced electromagnetic showers, it is relevant to explore lower ranges of values for $dE/dx$, since, as explained in Section~\ref{sec:erec}, a considerable amount of electromagnetic shower segments have a low $dE/dx$. The angular dependence for the different models is shown on the left side of  Figure~\ref{fig:dqdx_ratio} for several values for $dE/dx$.
The Birks model and the columnar modified box model have similar behaviors at higher $dE/dx$, but deviate significantly from one another for $dE/dx = 1.5$ MeV/cm. In fact, for both the ellipsoidal and columnar box models, for $dE/dx = 1.5$~MeV/cm, the ratio surpasses the value of 1 for some values of $\phi$. For both forms of the angular functions considered $f(90^\circ) = 1$ which is the minimum and
$f(0^\circ)$ is the maximal value. The fact that the ratio is surpassing 1 is unexpected as it would be implying that more charge is surviving recombination for smaller angles than at $\phi = 90^\circ$, for which the recombination factor is minimal.

To examine this more closely, in the right side of Figure \ref{fig:dqdx_ratio} is plotted the ratio between $dQ/dx(\phi)$ and $dQ/dx(90^\circ)$ as a function of the $dE/dx$, for the three models considered and for $\phi = 5^\circ, 25^\circ, 45^\circ, 85^\circ$. In the case of the Birks model, the ratio never quite reaches 1 but tends closer to it for the larger angles. On the other hand, in the case of the modified box models, for any $\phi$ value and in the range of $dE/dx < 2$ MeV/cm, the ratio goes above 1. This particular aspect of the prediction appears to be unphysical in a range that, while unconstrained by the current data, is important when considering electrons.  

As discussed in Ref.~\cite{icarus_angular}, the ellipsoidal modified box model was the one that best fit the ICARUS data. Both the angular-dependent Birks model and the columnar modified box model somewhat exaggerate the recombination effect for smaller values of $\phi$. The parameter $R$ provides an extra degree of freedom, with which the ellipsoidal model can be better fitted to the experimental data. 

The parameters $\alpha$, $\beta_{90}$ and $R$ for the ellipsoidal modified box model were obtained for values of $dE/dx$ larger than $5$~MeV/cm~\cite{icarus_angular}. However, as discussed previously, a significant amount of segments of the electromagnetic showers have $dE/dx$ in the range of 1 to 3~MeV/cm, where the model appears to have a nonphysical behavior. \textcolor{black}{Besides this, proton data is being used for electrons and we encounter this unphysical behaviour which might suggest that these values for the recombination model parameters or the shape of $f(\phi)$ are not valid for electrons and thus the parameters might not be particle independent.}
In order to ensure that recombination at any angle is never smaller than the effect at 90$^\circ$, thus avoiding the unphysical behaviour of the model, the model needs to be adapted for the $dE/dx$ range unconstrained by data. 

\begin{figure}[htbp]
\centering
\includegraphics[width=.6\textwidth]{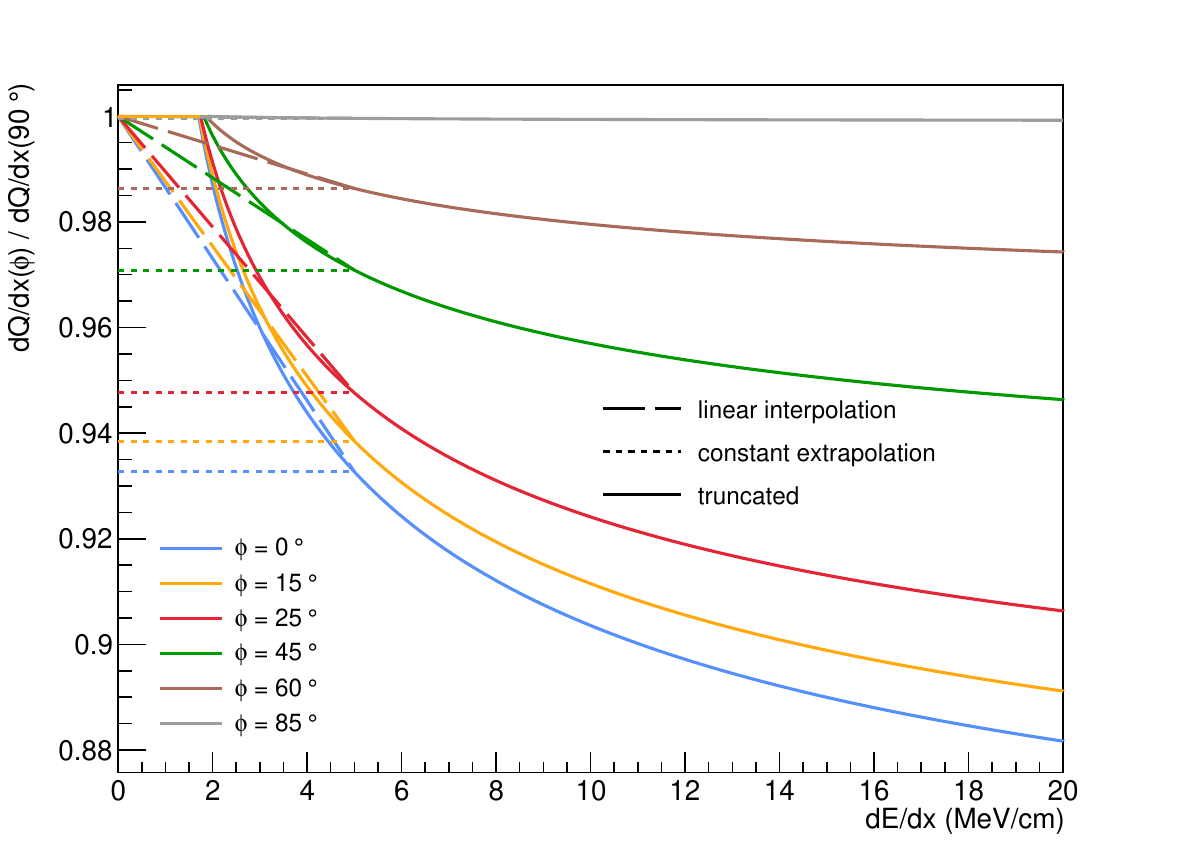}
\caption{Ratio between $dQ/dx (\phi)$ and $dQ/dx(90^{\circ})$ as a function of $dE/dx$ calculated considering the modified box model with an ellipsoidal angular dependence. For $dE/dx < 5$ Mev/cm, two corrections to the model were applied: a linear interpolation, corresponding to the dashed lines, a constant extrapolation, corresponding to the dotted lines, \textcolor{black}{and a scenario where the ratio is truncated at 1, corresponding to the full lines.} \label{fig:dqdx-correct}}
\end{figure}

Since the parameters of the model were obtained for values of $dE/dx$ larger than 5 MeV/cm, a correction is applied to the values below that threshold. In one scenario, \textcolor{black}{we required that the ratio must be 1 in the limit of $dE/dx=0$ MeV/cm, since there's no effect at any angle then.} A linear interpolation is made, where a straight line is calculated from $\frac{dQ/dx(\phi)}{dQ/dx(90^\circ)} = 1$ to the value of the ratio predicted by the model for $dE/dx = 5$ MeV/cm, as shown in the dashed lines in Figure~\ref{fig:dqdx-correct}.
\textcolor{black}{This is a very simple first approach to the problem that satisfies the constraints.}

\textcolor{black}{We considered two additional extreme scenarios in order to bound the range of the effect, given the lack of constraining data at low dE/dx. In the second scenario, the ratio} is extrapolated at a constant value for $dE/dx < 5$ MeV/cm. This is an extreme scenario \textcolor{black}{with no physical justification}, only considered here \textcolor{black}{provide} an upper limit to the angular dependent recombination effect. \textcolor{black}{The only constraint satisfied in this scenario is that the ratio for $dE/dx < 5$ MeV/cm should not be smaller than the ratio for $dE/dx = 5$ MeV/cm.} This constant extrapolation scenario is shown in the dotted lines of Figure~\ref{fig:dqdx-correct}.

\textcolor{black}{In the third scenario considered, the ellipsoidal model was applied bellow the 5 MeV/cm threshold but for the values of $dE/dx$ where the ratio exceeds 1, the ratio is truncated at 1. This scenario provides a lower limit to the angle-dependent recombination effect and is shown in the full lines of Figure~\ref{fig:dqdx-correct}.}

\section{Impact on neutrino energy reconstruction}
\label{sec:impact}
With the ellipsoidal modified box model and the corrections described previously, we can now describe the impact of recombination on the neutrino energy reconstruction by studying the effect in electromagnetic showers. 

Using the second data set of simulated electromagnetic showers described in Section~\ref{sec:data}, the direction of each segment of the shower as well as the deposited energy can be obtained. With this, the modified box model was used to calculate the $dQ/dx$, with and without considering the ellipsoidal angular dependence.

\begin{figure}[htbp]
\centering
\includegraphics[width=.45\textwidth]{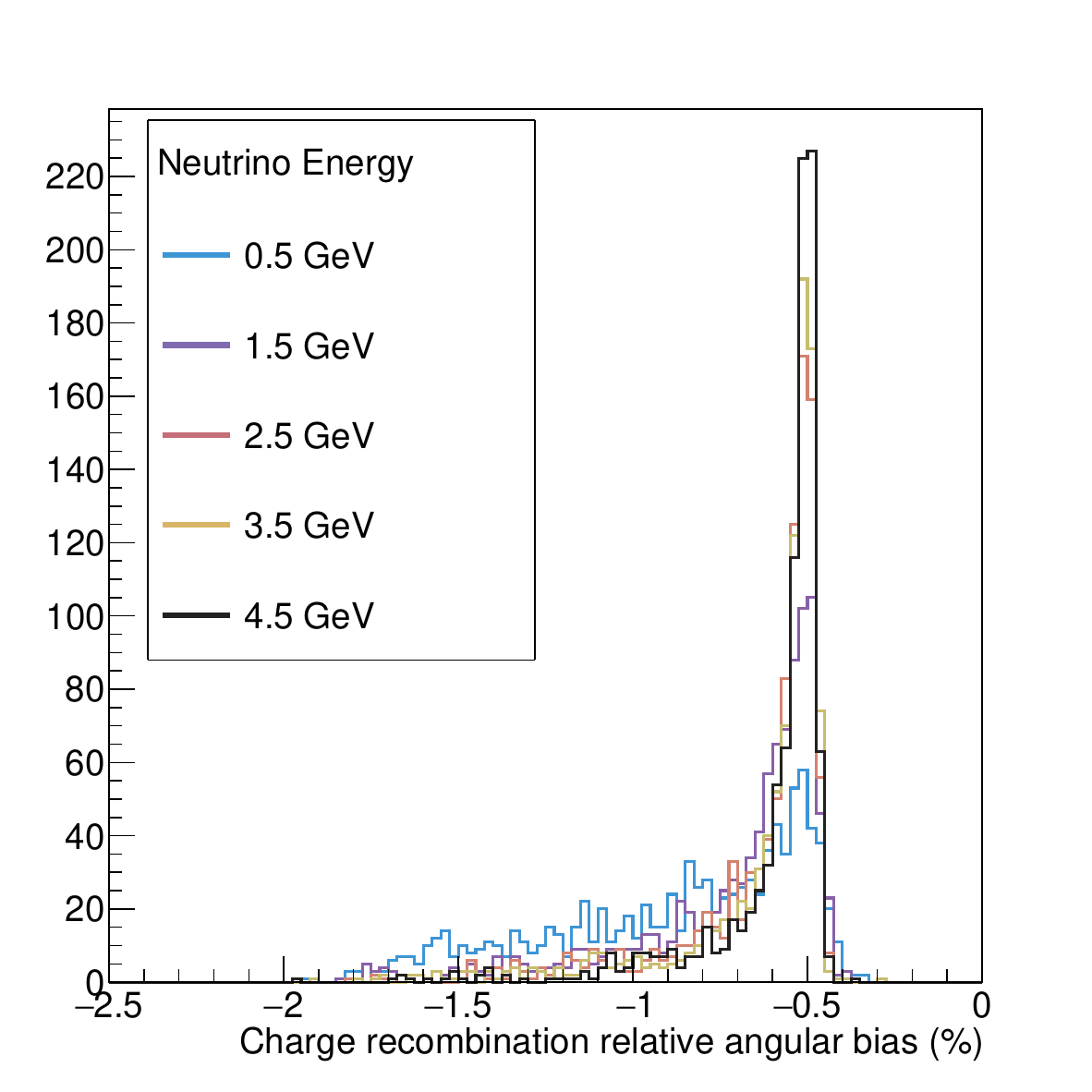}
\qquad
\includegraphics[width=.45\textwidth]{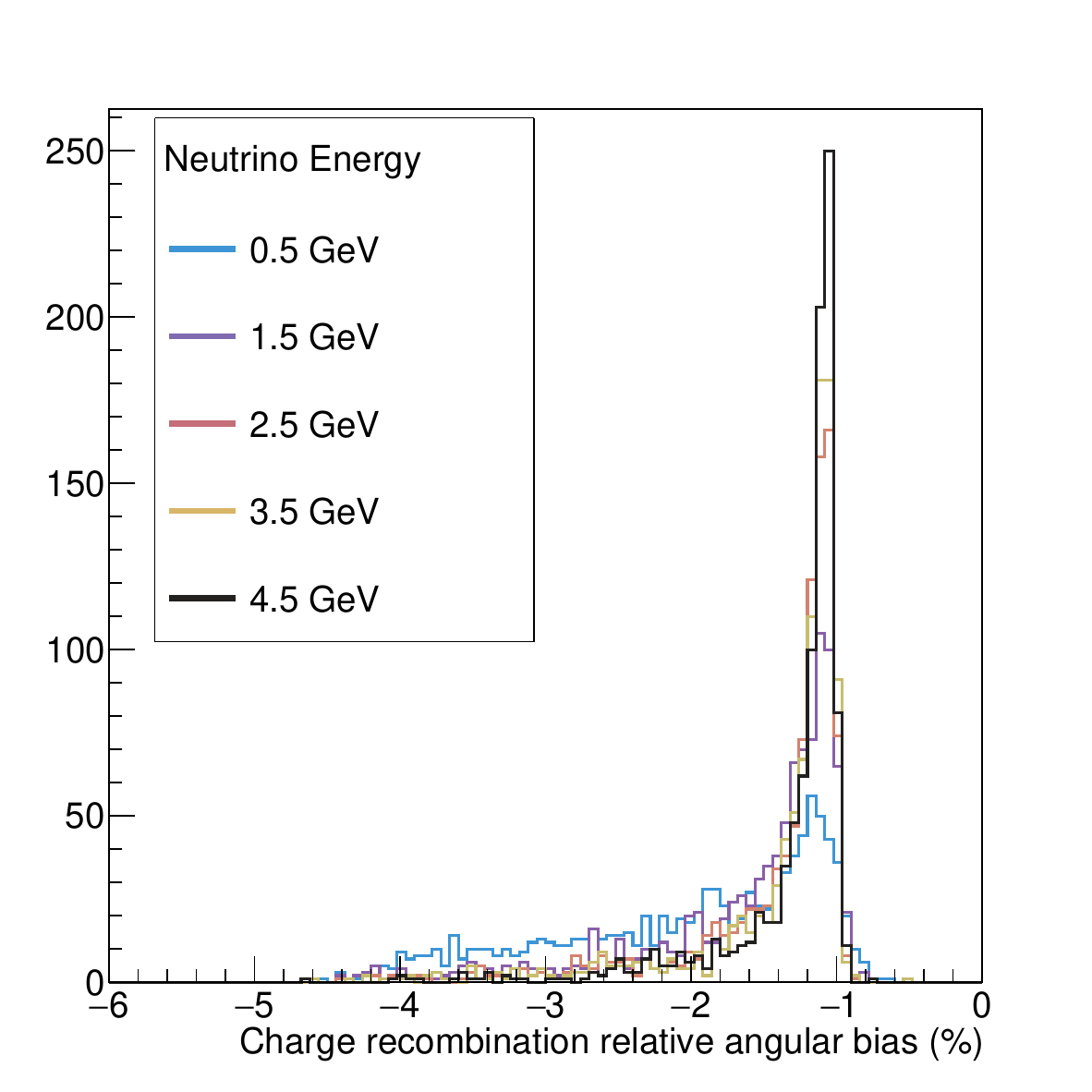}
\includegraphics[width=.45\textwidth]{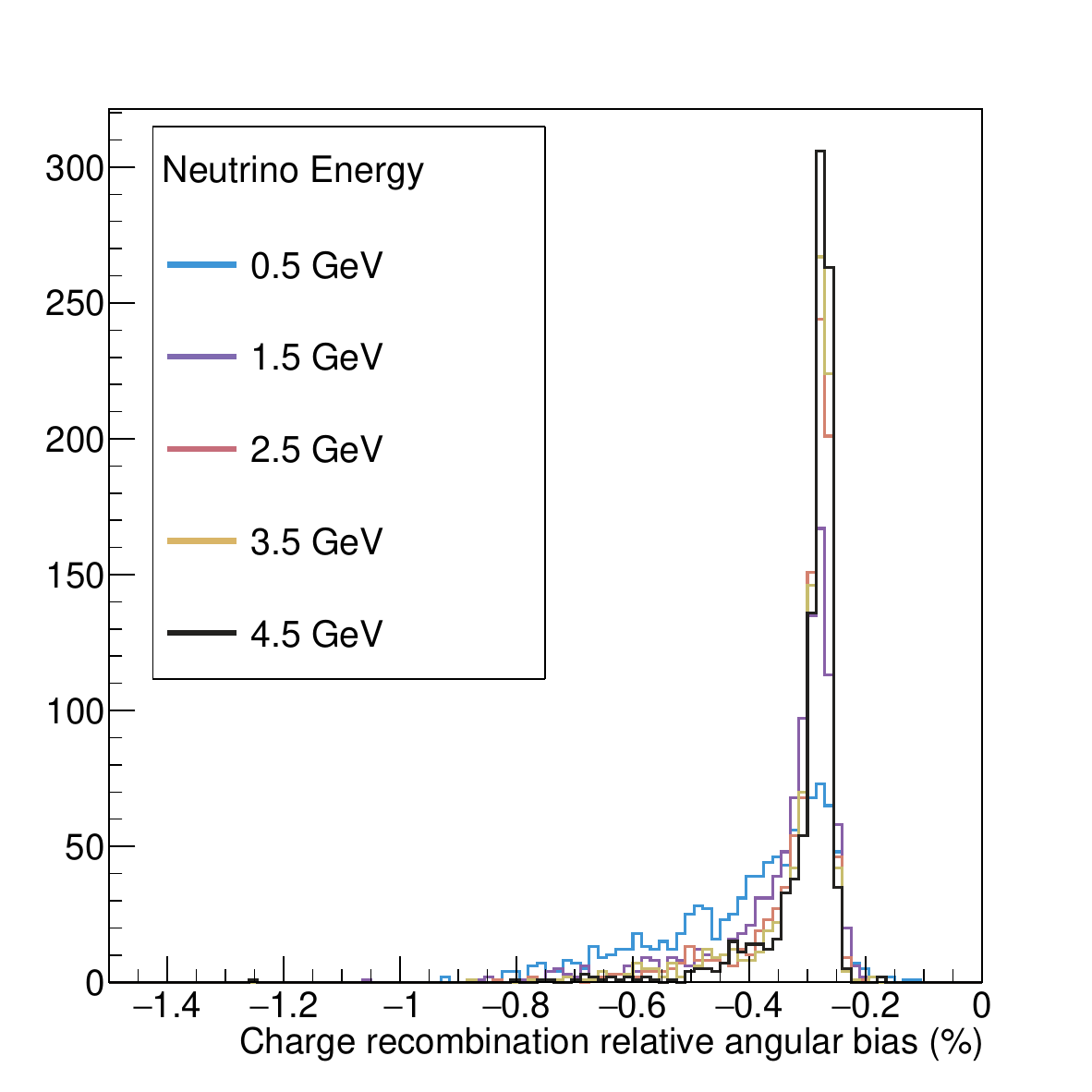}
\caption{Distribution of the recombination relative angular bias for different values for the neutrino energy. For the calculations of the collected charge, the ellipsoidal modified box model was used, considering the linear interpolation (left), the constant extrapolation (right)\textcolor{black}{, and the truncated scenario (bottom), following the} corrections described in the text.\label{fig:dqr}}
\end{figure}

The impact of the angular dependence of the recombination effect on collected charge of electromagnetic showers can be described by the charge recombination relative angular bias:
\begin{equation}
    \label{eq:5.1}
    \frac{Q' - Q}{Q}
\end{equation}
where $Q'$ is calculated by integrating equation \eqref{eq:4.4} for all track segments in the shower, considering an angular dependence and $Q$ is calculated by integrating equation \eqref{eq:4.4} without the angular dependence, by setting $f(\phi)=1$. This bias translates how much the collected charge of the shower is reduced when considering the angular dependence of the recombination model. In Figure \ref{fig:dqr} are shown the distributions of the relative angular bias using the two considered corrections to the ellipsoidal modified box model for several values of neutrino energy. 

 Since the angular distribution of the electrons is clustered around $90^\circ$, especially for high energies, the distributions of the relative angular bias show prominent peaks. In the case of the linear interpolated correction, the peaks of the distributions are around $-0.6\%$. As for the constant extrapolated correction, the bias is larger with the peaks around $-1.1\%$. \textcolor{black}{The extreme case where the ratio was truncated, is where the bias is the smallest, with peaks around $-0.3\%$.} However, the relative angular bias is always significantly smaller than 0, which comes from the fact that we not only take into consideration the angle between the initial electron and the electric field but also the angle between each shower segment and the electric field. The left tail is more pronounced \textcolor{black}{and so the bias is more significant} for smaller energies since $\phi$ is more widely distributed in these energy ranges.

\begin{figure}[htbp]
\centering
\includegraphics[width=.6\textwidth]{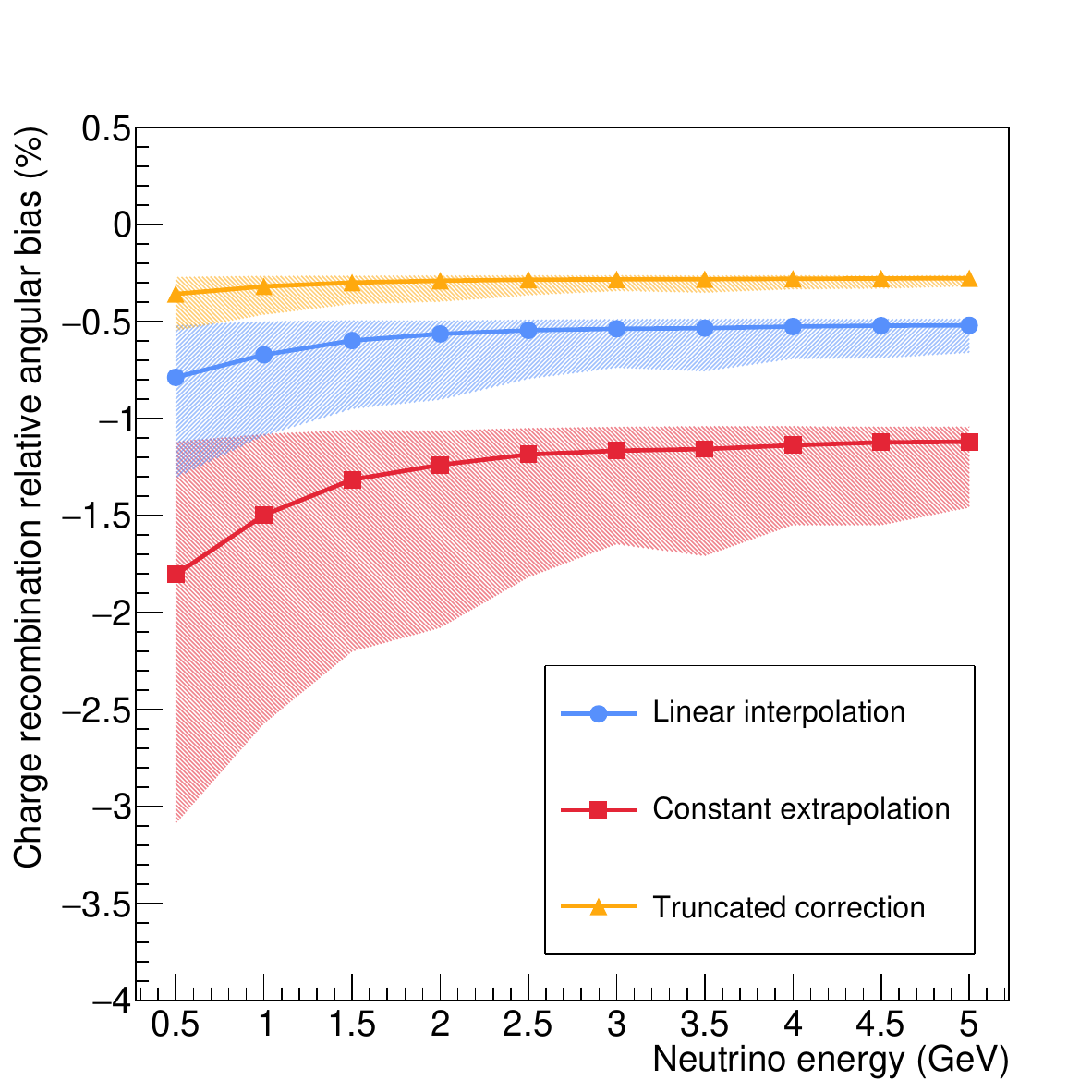}
\caption{Median values of the charge recombination relative angular bias as a function of the neutrino energy. The shaded areas correspond to the region that contains $68.3\% (\pm 1 \sigma)$ of each distribution. \label{median_bias}}
\end{figure}

From these distributions, the median values of the relative angular bias were calculated and are plotted in Figure \ref{median_bias}. 
The shaded areas correspond to the regions of values that contain $68.3\% (\pm 1 \sigma)$ of the distribution. With both the interpolated and extrapolated corrections, the median value of the relative angular bias varies considerably with the neutrino energy. In the first case, the bias varies from $-0.8\%$ for a $0.5$ GeV neutrino to $-0.6\%$ for a $5$ GeV neutrino. As for the extreme scenario, the bias varies from $-1.8\%$ for a $0.5$ GeV neutrino to $-1.2\%$ for a $5$ GeV neutrino. In the lower energy range of neutrino energy, the reduction of the collected charge of the shower is quite significant.  \textcolor{black}{For the third scenario, the bias varies from $-0.35\%$ for a $0.5$ GeV neutrino to $-0.28\%$ for a $5$ GeV neutrino, representing a much smaller dependence on the neutrino energy than the other two cases.}

The variation of the relative angular bias with the neutrino energy is not unexpected as explained previously. For lower energy neutrinos, $\phi$ is uniformly distributed, including smaller angles for which recombination is stronger. In the case of higher energy neutrinos, the distribution of $\phi$ is centered around $90^\circ$ for which the recombination effect is minimal. On the other hand, the fact that the relative angular bias does vary with the neutrino energy can pose a challenge when attempting to correct it.

\textcolor{black}{It should be noted the values for the median between the three cases represent a systematic uncertainty band, where the true value of bias must lie and that the values for the relative angular bias depend strongly on the correction applied to the model.}

\textcolor{black}{
In the future, it would be relevant to repeat this study for pions, since as seen in Section \ref{sec:erec}, the distribution of the pion angles also varies with the neutrino energy and they have a larger $dE/dx$ than that of electrons. Pions are produced at higher energies, where the relative angular bias should be smaller and thus, the effect should not be as prominent as with electrons. Still, it would be interesting to consider the effects in both leptonic and hadronic components in order to arrive at the final neutrino energy bias. The angular bias on the final neutrino energy could be significant enough to affect oscillation analyses.}

\textcolor{black}{
Calibrating this effect with energy response functions will be challenging due the the energy- dependence of the effect. We believe it will be necessary to constrain better the microscopic treatment of the angular dependence of the recombination process. It may be possible to carry this out with stopping cosmic ray muons, especially in a geometry with vertical electric field, in which a larger fraction of muons the muons will have  small angles with respect to the field.}

\section{Conclusion}
\label{sec:end}

In this paper we analyzed the impact of recombination models with electric field angle dependence on the calorimetric reconstruction of electron neutrino energy in LArTPCs. In order to avoid problems in the predictions at high energy losses present in the Birks and simple box model, most of the latest empirical measurements focus primarily on the modified box model and its extension to include an angular-dependence, the ellipsoidal model.
We verified that the angular-dependence of the modified box model is also problematic in the low energy loss range typical of minimum ionizing particles. This region is in fact largely unconstrained by data, so we considered simple (but physically motivated) corrections in order to allow our study for the response to electrons with angles and energies resulting from CC interactions by neutrinos with a direction normal to the electric field and energy in the range of 0.5 to 5 GeV.

We concluded that an energy bias between -0.9\% at 0.5 GeV and -0.6\% at 5 GeV in calorimetric electron neutrino energy can arise between a model with and without recombination angular dependence. These values, while still contained within the uncertainty budget of future experiments, are large enough to justify efforts to measure this effect at low $dE/dx$ and taking it into account in order to achieve the best precision in future LArTPC electron neutrino calorimetry.

\acknowledgments

The authors would like to thank Wei Shi and Callum Wilkinson for kindly providing the simulation data used in this analysis. We also thank Chris Marshall for useful conversations. 
The authors acknowledge the support of Fundação para a Ciência e a Tecnologia (Portugal) through grants 2024.00258.CERN and CEECINST/00032/2021/CP2790/CT0001,

\bibliographystyle{JHEP}
\bibliography{biblio.bib}

\providecommand{\href}[2]{#2}\begingroup\raggedright\begin{thebibliography}{10}

\bibitem{Rubbia:1977zz}
C.~Rubbia, \emph{{The Liquid Argon Time Projection Chamber: A New Concept for Neutrino Detectors}},  Tech. Rep. CERN-EP-INT-77-08, CERN-EP-77-08, CERN (5, 1977).

\bibitem{ICARUS:2004wqc}
{\scshape ICARUS} collaboration, \emph{{Design, construction and tests of the ICARUS T600 detector}}, \href{https://doi.org/10.1016/j.nima.2004.02.044}{\emph{Nucl. Instrum. Meth. A} {\bfseries 527} (2004) 329}.

\bibitem{MicroBooNE:2016pwy}
{\scshape MicroBooNE} collaboration, \emph{{Design and Construction of the MicroBooNE Detector}}, \href{https://doi.org/10.1088/1748-0221/12/02/P02017}{\emph{JINST} {\bfseries 12} (2017) P02017} [\href{https://arxiv.org/abs/1612.05824}{{\ttfamily 1612.05824}}].

\bibitem{SBND:2025lha}
{\scshape SBND} collaboration, \emph{{The Short-Baseline Near Detector at Fermilab}},  Tech. Rep. FERMILAB-PUB-25-0154-PPD, Fermilab (3, 2025).

\bibitem{DUNE:2020lwj}
{\scshape DUNE} collaboration, \emph{{Deep Underground Neutrino Experiment (DUNE), Far Detector Technical Design Report, Volume I Introduction to DUNE}}, \href{https://doi.org/10.1088/1748-0221/15/08/T08008}{\emph{JINST} {\bfseries 15} (2020) T08008} [\href{https://arxiv.org/abs/2002.02967}{{\ttfamily 2002.02967}}].

\bibitem{MicroBooNE:2019efx}
{\scshape MicroBooNE} collaboration, \emph{{Calibration of the charge and energy loss per unit length of the MicroBooNE liquid argon time projection chamber using muons and protons}}, \href{https://doi.org/10.1088/1748-0221/15/03/P03022}{\emph{JINST} {\bfseries 15} (2020) P03022} [\href{https://arxiv.org/abs/1907.11736}{{\ttfamily 1907.11736}}].

\bibitem{DUNElbl}
{\scshape DUNE} collaboration, \emph{{Long-baseline neutrino oscillation physics potential of the DUNE experiment}}, \href{https://doi.org/10.1140/epjc/s10052-020-08456-z}{\emph{Eur. Phys. J. C} {\bfseries 80} (2020) 978} [\href{https://arxiv.org/abs/2006.16043}{{\ttfamily 2006.16043}}].

\bibitem{jaffe}
G.~Jaffé, \emph{{Zur Theorie der Ionisation in Kolonnen}}, \href{https://doi.org/https://doi.org/10.1002/andp.19133471205}{\emph{Annalen der Physik} {\bfseries 347} (1913) 303}.

\bibitem{onsager}
L.~Onsager, \emph{{Initial Recombination of Ions}}, \href{https://doi.org/10.1103/PhysRev.54.554}{\emph{Phys. Rev.} {\bfseries 54} (1938) 554}.

\bibitem{box}
J.~Thomas and D.A.~Imel, \emph{{Recombination of electron-ion pairs in liquid argon and liquid xenon}}, \href{https://doi.org/10.1103/PhysRevA.36.614}{\emph{Phys. Rev. A} {\bfseries 36} (1987) 614}.

\bibitem{icarus_recom}
{\scshape ICARUS} collaboration, \emph{{Study of electron recombination in liquid argon with the ICARUS TPC}}, \href{https://doi.org/10.1016/j.nima.2003.11.423}{\emph{Nucl. Instrum. Meth. A} {\bfseries 523} (2004) 275}.

\bibitem{argoneut_recom}
{\scshape ArgoNeuT} collaboration, \emph{{A Study of Electron Recombination Using Highly Ionizing Particles in the ArgoNeuT Liquid Argon TPC}}, \href{https://doi.org/10.1088/1748-0221/8/08/P08005}{\emph{JINST} {\bfseries 8} (2013) P08005} [\href{https://arxiv.org/abs/1306.1712}{{\ttfamily 1306.1712}}].

\bibitem{icarus_angular}
{\scshape ICARUS} collaboration, \emph{{Angular dependent measurement of electron-ion recombination in liquid argon for ionization calorimetry in the ICARUS liquid argon time projection chamber}}, \href{https://doi.org/10.1088/1748-0221/20/01/P01033}{\emph{JINST} {\bfseries 20} (2025) P01033} [\href{https://arxiv.org/abs/2407.12969}{{\ttfamily 2407.12969}}].

\bibitem{GENIE}
C.~Andreopoulos et~al., \emph{{The GENIE Neutrino Monte Carlo Generator}}, \href{https://doi.org/10.1016/j.nima.2009.12.009}{\emph{Nucl. Instrum. Meth.} {\bfseries A614} (2010) 87} [\href{https://arxiv.org/abs/0905.2517}{{\ttfamily 0905.2517}}].

\bibitem{Stowell:2016jfr}
P.~Stowell et~al., \emph{{NUISANCE: a neutrino cross-section generator tuning and comparison framework}}, \href{https://doi.org/10.1088/1748-0221/12/01/P01016}{\emph{JINST} {\bfseries 12} (2017) P01016} [\href{https://arxiv.org/abs/1612.07393}{{\ttfamily 1612.07393}}].

\bibitem{edep-sim}
C.~McGrew, ``{edep-sim : An Energy Deposition Simulation}.'' \url{https://github.com/ClarkMcGrew/edep-sim}, 2021.

\bibitem{ALLISON2016186}
J.~Allison, K.~Amako, J.~Apostolakis, P.~Arce, M.~Asai, T.~Aso et~al., \emph{{Recent developments in Geant4}}, \href{https://doi.org/https://doi.org/10.1016/j.nima.2016.06.125}{\emph{Nuclear Instruments and Methods in Physics Research Section A: Accelerators, Spectrometers, Detectors and Associated Equipment} {\bfseries 835} (2016) 186}.

\bibitem{1610988}
J.~Allison, K.~Amako, J.~Apostolakis, H.~Araujo, P.~Arce~Dubois, M.~Asai et~al., \emph{{Geant4 developments and applications}}, \href{https://doi.org/10.1109/TNS.2006.869826}{\emph{IEEE Transactions on Nuclear Science} {\bfseries 53} (2006) 270}.

\bibitem{AGOSTINELLI2003250}
{\scshape GEANT4} collaboration, \emph{{GEANT4 - A Simulation Toolkit}}, \href{https://doi.org/10.1016/S0168-9002(03)01368-8}{\emph{Nucl. Instrum. Meth. A} {\bfseries 506} (2003) 250}.

\bibitem{NIST}
M.~Berger et~al., \emph{{ESTAR, PSTAR, and ASTAR: Computer Programs for Calculating Stopping-Power and Range Tables for Electrons, Protons, and Helium Ions (version 1.2.3). [Online]}}, \href{https://doi.org/https://dx.doi.org/10.18434/T4NC7P}{\emph{National Institute of Standards and Technology, Gaithersburg, MD} (2025) }.

\bibitem{DUNE:2021tad}
{\scshape DUNE} collaboration, \emph{{Deep Underground Neutrino Experiment (DUNE) Near Detector Conceptual Design Report}}, \href{https://doi.org/10.3390/instruments5040031}{\emph{Instruments} {\bfseries 5} (2021) 31} [\href{https://arxiv.org/abs/2103.13910}{{\ttfamily 2103.13910}}].

\bibitem{lar-RECOM-prop}
E.~Segreto, \emph{{Properties of charge recombination in liquid argon}}, \href{https://doi.org/10.1103/PhysRevD.110.062002}{\emph{Phys. Rev. D} {\bfseries 110} (2024) 062002} [\href{https://arxiv.org/abs/2405.00905}{{\ttfamily 2405.00905}}].

\bibitem{ellipsoidal}
V.~Cataudella, A.~de~Candia, G.D.~Filippis, S.~Catalanotti, M.~Cadeddu, M.~Lissia et~al., \emph{{Directional modulation of electron-ion pairs recombination in liquid argon}}, \href{https://doi.org/10.1088/1748-0221/12/12/P12002}{\emph{JINST} {\bfseries 12} (2017) P12002}.

\end{thebibliography}\endgroup

\end{document}